\documentclass{aa}  
\usepackage{graphicx}
\usepackage{subfigure}
\usepackage{txfonts}
\usepackage{verbatim}
\usepackage{float}
\usepackage{pifont}
\usepackage{tabularx,adjustbox,booktabs}
\begin{document} 

\title{Asteroseismology of ZZ Ceti stars with full evolutionary white
  dwarf models. } \subtitle{II. The impact of AGB thermal pulses on the asteroseismic inferences of ZZ Ceti stars.}

\author{F. C. De Ger\'onimo\inst{1,2}, L. G. Althaus\inst{1,2},
  A. H. C\'orsico\inst{1,2}, A. D. Romero\inst{3} and S. O. Kepler\inst{3}
}

\institute{$^{1}$ Grupo de Evoluci\'on Estelar y Pulsaciones. Facultad
  de Ciencias Astron\'omicas y Geof\'{\i}sicas, Universidad Nacional
  de La Plata, Paseo del Bosque s/n, 1900 La Plata, Argentina\\ 
  $^{2}$ Instituto de Astrof\'isica La Plata (IALP - CONICET)\\ 
  $^{3}$ Departamento de Astronomia, Universidade
  Federal do Rio Grande do Sul, Av. Bento Goncalves 9500, Porto Alegre
  91501-970, RS, Brazil\\ 
 \email{fdegeronimo,althaus,acorsico@fcaglp.unlp.edu.ar,alejandra.romero@ufrgs.br,kepler@if.ufrgs.br} }
\date{Received ; accepted }

\abstract{The thermally pulsing phase on the asymptotic giant branch
  (TP-AGB) is the last nuclear burning phase experienced by most of
  low and intermediate mass stars. During this phase, the outer
  chemical stratification above the C/O core of the emerging white
  dwarf is built up. The chemical structure resulting from progenitor
  evolution strongly impacts the whole pulsation spectrum exhibited by
  ZZ Ceti stars, which are pulsating C/O core white dwarfs located on
  an narrow instability strip at  $T_{\rm eff} \sim 12000$
  K. 
  Several physical processes occurring during
  progenitor evolution strongly affect the chemical structure of these
  stars, being those found during the TP-AGB phase ones of the most relevant for the pulsational properties of ZZ Ceti stars.} 
  {We present a
  study of the impact of the chemical structure built up during the TP-AGB evolution on   the stellar parameters inferred from asteroseismological fits of ZZ Ceti stars.}  {Our analysis is based on a set
  of carbon-oxygen core white dwarf models with masses from 0.534 to
  $0.6463 M_{\sun}$ derived from full evolutionary computations from
  the ZAMS to the ZZ Ceti domain.  We compute evolutionary sequences
  that experience different number of thermal pulses.
}{ 
  We find that the occurrence or not of thermal pulses during AGB evolution implies an average deviation 
  in the asteroseimological effective temperature of ZZ Ceti  stars of at most
  8\% and of the order of $\lesssim$ 5\% in the stellar mass. For the mass
  of the hydrogen envelope, however, we find deviations up to 2 orders of magnitude in the case of cool ZZ Ceti  stars. For hot and intermediate temperature ZZ Ceti  stars shows no differences in the hydrogen envelope mass in most cases.}{Our results show that, in general, the impact of the occurrence or not of thermal pulses in the progenitor stars is not negligible and must be taken into account  in asteroseismological studies of ZZ Ceti stars.}

\titlerunning{Pulsating white dwarfs}
\maketitle

\section{Introduction}
\label{introduction}

After the end of core He burning, single low- and intermediate-mass
stars --up to 8-10 $M_{\odot}$-- experience the last nuclear burning
phase in their evolution, known as the thermal pulses (TP), while
ascending along the asymptotic giant branch (AGB). Stars at this
evolutionary stage are composed by a C/O core, product of the core He
burning stage, surrounded by a He burning shell, and surrounded itself
by a He buffer and an outer H-burning shell.  As the star begins to
ascend to the AGB, the H-burning layer is turned off.  After its
reignition, the He shell becomes unstable and the star begins the
thermally pulsing stage (TP-AGB).

The interplay between mixing and burning that occurs along the TP-AGB
leaves marked imprints on the chemical stratification of the outer
layers of the emerging white dwarf (WD) \citep{2010ApJ...717..897A}.
Indeed, during the TP-AGB, third dredge-up episodes yield appreciable surface 
composition changes, as a result of which the
star becomes C enriched. In addition, some
stars may experience hot bottom burning (HBB) episodes where the
bottom of the convective envelope penetrates into the top of the
H-burning shell producing N, Na and Al
\citep{2014PASA...31...30K}. 
Most importantly for the inner chemical stratification that will characterize the emerging WD
is the formation of an intershell region rich in He and C during the TP-AGB phase that is left by the 
short-lived He flash induced convection zone at the peak flash. The mass of this intershell
(and the position of the core-He transition) 
depends strongly on both the occurrence of overshooting (OV) during the He flash and the number of thermal 
pulses (TP). The latter is determined by the initial mass, composition and by the poorly
constrained efficiency of mass loss rate \citep{2014PASA...31...30K}. 
It should be mentioned that, despite the relevance of the occurrence of TPs for
the chemical structure of the outer layers of WDs, their consequences for the
asteroseismological fits of ZZ Ceti stars are not usually taken into account, and have not been assessed.

Although the TP-AGB phase is expected for most of WD
progenitors, it is possible that some WDs could have resulted
from progenitor stars that avoided this phase.  
For instance, it is well known that low-mass He-burning stars located
at the extreme horizontal branch ---and thus characterized by
extremely thin H envelopes--- are expected to evolve directly to the
WD stage avoiding the AGB, the so called AGB-Manqué and post early AGB
stars \citep{1990ApJ...364...35G}. In line with this, recent evidence
suggest that most He rich stars of NGC2808 do not reach the AGB phase,
evolving directly to the WD state after the end of the He core burning
\citep{2017arXiv170602278M}. In addition, departure from the AGB
before reaching the TP-AGB phase as a result of mass transfer by
binary interaction \citep{2006ASPC..353..149P,2000MNRAS.319..215H} or
envelope ejection by the swallowing of a planet or a very low mass
companion \citep{2002PASP..114..602D} is also possible.

In view of these considerations, we must consider that a fraction of WDs have evolved from progenitors that avoided the
TP-AGB phase. This being the case, differences in the chemical
structure of the outermost layers of the C/O core should be expected
according to whether the progenitor stars evolved through the TP-AGB
or not, with consequences for the expected pulsational properties of
pulsating WDs.

In the context of the preceeding discussion, it is well known that the
precise shape of the chemical abundance distribution is critical
for the pulsational properties of H-rich WDs (ZZ Ceti stars).  ZZ
Ceti (or DAV) variable stars are located in an instability strip with
effective temperatures between 10500 K and 12500 K
\citep{2008PASP..120.1043F,2008ARA&A..46..157W,2010A&ARv..18..471A,2017EPJWC.15201011K},
and constitute the most numerous class of compact pulsators. These
stars are characterized by multimode photometric variations caused by
non-radial {\it g}-mode pulsations of low harmonic degree ($\ell \leq
2$) with periods between 70 and 1500 s and amplitudes up to 0.30 mag.
These pulsations are thought to be excited by both the $\kappa-\gamma$
mechanism acting on the base of the H partial ionization zone
\citep{1981A&A...102..375D,1982ApJ...252L..65W} and the ``convective
driving'' mechanism
\citep{1991MNRAS.251..673B,1999ApJ...511..904G,2013EPJWC..4305005S}.
Since the discovery of the first ZZ Ceti star, HL Tau 76 by
\citet{1968ApJ...153..151L}, a continuous effort has been made to
model the interior of these variable stars.

From the asteroseismological analysis of DAVs, i.e. the
comparison of the observed periods in variable  DA WDs with those computed
from theoretical models, details about the previous
evolutionary history of the star can be inferred.
This analysis allow us to constrain several stellar parameters such
as the stellar mass, the thickness of the outer envelopes, the core
chemical composition, and the stellar rotation rates
\citep[e.g.,][]{2012MNRAS.420.1462R}.  In addition, ZZ Ceti
asteroseismology is a valuable tool for studying axions
\citep{Isern92,Isern10,2008ApJ...675.1505B,
  2012MNRAS.424.2792C,2016JCAP...07..036C} and crystallization
\citep{1999ApJ...525..482M,2004A&A...427..923C,2005A&A...429..277C,
  2004ApJ...605L.133M,2005A&A...432..219K,2013ApJ...779...58R}.

Two main approaches have been adopted for WD asteroseismology. The
first one employs static stellar models with parametrized chemical
profiles.  This approach has the advantage that it allows a full
exploration of the parameter space to find an optimal seismic model
\citep{1998ApJS..116..307B,2001ApJ...552..326B,2009MNRAS.396.1709C,2011ApJ...742L..16B,2014ApJ...794...39B}, 
  ultimately leading to good matches to the periods
  \citep{2017A&A...598A.109G,2017ApJ...834..136G}. 
 As precise as they have become, parameterized approaches rely on
  an educated guess of the internal composition profiles, due to the
  large number of parameters involved and the small number of periods
  typically available. Such profiles come from fully evolutionary
  models, and this is the approach we follow here. The models are
  generated by tracking the complete evolution of the progenitor star,
  from the Zero Age Main Sequence (ZAMS) to the WD stage
  \citep{2012MNRAS.420.1462R,2013ApJ...779...58R}.  This approach
involves the most detailed and updated input physics, in particular
regarding the internal chemical structure expected from the nuclear
burning history of the progenitor, a crucial aspect for correctly
disentangle the information encoded in the pulsation patterns of
variable WDs. This asteroseismic approach allows not only to unveil
the interior structures of stars, but also find out how a star gets to
that structure.  This method has been successfully applied in
different classes of pulsating WDs \citep[see][in the case of ZZ Ceti,
  DB and PG1159 stars,
  respectively]{2012MNRAS.420.1462R,2013ApJ...779...58R,2006A&A...454..863C,2006A&A...458..259C,2009A&A...506..835C}.

However, none of the asteroseismological approaches take into account
the current uncertainties in stellar evolution, neither in the
modeling nor in the input physics of WD progenitors. In this regard,
\citet{2017A&A...599A..21D} explored for the first time the impact of
the occurrence of TP in WD progenitors, the uncertainty in the
$\rm{}^{12}C(\alpha,\gamma)^{16}O$ cross section and the ocurrence of
extra mixing on the expected period spectrum of ZZ Ceti stars. They
reported that the chemical profiles built up during the TP-AGB phase
impact markedly the $g$-mode pulsational periods, and concluded that
the occurrence or not of the TP-AGB phase constitutes a relevant issue
that has to be taken into account in seismological fits of these
stars.

In view of the findings by \citet{2017A&A...599A..21D}, the present
paper is focused on assessing the role played by the TP-AGB phase of 
progenitor stars in the  stellar parameters inferred from
asteroseismological fits to artificial and real ZZ Ceti
stars. 
To do this we first computed evolutionary sequences from the ZAMS to
the TP-AGB stage. 
During the AGB, we forced the models
to leave this stage at two instances: previous to the first thermal
pulse and at the end of the third thermal pulse, thus generating two sets
of evolutionary models. Next, we followed the evolution of the progenitors to
the WD state until the domain of the ZZ Ceti instability strip, where
we computed the period spectrum. Finally, we performed the
asteroseismological analysis to sets of random periods representative
of ZZ Ceti stars and later to real stars by considering the two sets of
models.

This work is part of an ongoing project in which we will assess  uncertainties in evolutionary history of WD progenitors and their impact on the 
 pulsational
properties and asteroseismological fits to ZZ Ceti stars.  The work is
organized as follow: in Sect. \ref{cap:tools} we introduce the
numerical tools employed and the input physics assumed in the
evolutionary calculations together with the pulsation code employed.
In Sect.\ref{cap:models-asteroseismological} we present our results
and, finally, in Sect. \ref{cap:conclusions} we conclude the paper by
summarizing our findings.

\section{Computational tools}
\label{cap:tools}

\subsection{Evolutionary code and input physics}

The DA WD evolutionary models computed in this work were generated
with the {\tt LPCODE} evolutionary code. {\tt LPCODE} produces
detailed WD models in a consistent way with the predictions of
progenitor history. The code is based on an updated physical
description \citep{2005A&A...435..631A,2010ApJ...717..897A,
  2010ApJ...717..183R,2012MNRAS.420.1462R,2016A&A...588A..25M} and was
employed to study different aspects of the evolution of low-mass stars
\citep{2011A&A...533A.139W, 2013A&A...557A..19A,2015A&A...576A...9A},
formation of horizontal branch stars \citep{2008A&A...491..253M},
extremely low mass WDs \citep{2013A&A...557A..19A}, AGB, and post-AGB
evolution \citep{2016A&A...588A..25M}. We enumerate below the most
important physical parameters relevant to this work: {\it i)} for
pre-WD stages we adopted the standard mixing-length theory with the
free parameter $\alpha = 1.61$; {\it ii)} diffusive  overshooting
during the evolutionary stages prior to the TP-AGB phase was allowed
to occur following the description of \citet{1997A&A...324L..81H}. We
adopted $f= 0.016$ for all sequences. The occurrence of OV is relevant
for the final chemical stratification of the WD
\citep{2002ApJ...581..585P,2003ApJ...583..878S}; {\it iii)} breathing
pulses, which occur at the end of core He burning, were suppressed
\citep[see][for a discussion on this topic]{2003ApJ...583..878S}; {\it
  iv)} a simultaneous treatment of non-instantaneous mixing and
burning of elements were considered. Our nuclear network accounts
explicitly for the following 16 elements: $^1$H, $^2$H, $^3$He,
$^4$He, $^7$Li, $^7$Be, $^{12}$C, $^{13}$C, $^{14}$N, $^{15}$N,
$^{16}$O, $^{17}$O, $^{18}$O, $^{19}$F, $^{20}$Ne, and $^{22}$Ne, and
34 thermonuclear reaction rates \citep{2005A&A...435..631A}; {\it v)}
gravitational settling and thermal and chemical diffusion were taken
into account during the WD stage for $^1$H, $^3$He,
$^4$He,$^{12}$C,$^{13}$C, $^{14}$N, and $^{16}$O
\citep{2003A&A...404..593A}; {\it vi)} during the WD phase, chemical
rehomogenization of the inner C-O profile induced by Rayleigh-Taylor
(RT) instabilities was implemented following
\citet{1997ApJ...486..413S}; {\it vii)} for the high-density regime
characteristic of WDs, we used the equation of state of
\citet{1994ApJ...434..641S}, which accounts for all the important
contributions for both the liquid and solid phases.
  
Our evolutionary sequences were computed considering all the
evolutionary stages of the WD progenitor, including the stable core He
burning and the TP-AGB and post-AGB phases.  It is worth mentioning
that {\tt LPCODE} has been compared against other WD evolution code,
showing $2 \%$ differences in the WD cooling times that come from the
different numerical implementations of the stellar evolution equations
\citep{2013A&A...555A..96S}.

\subsection{Pulsation code}
\label{pulsation-codes}

We employed the {\tt LP-PUL} adiabatic non-radial pulsation code
described in \citet{2006A&A...454..863C} for the pulsation analysis
presented in this work. {\tt LP-PUL} is coupled to the {\tt LPCODE}
evolutionary code and solves the full fourth-order set of real
equations and boundary conditions governing linear, adiabatic and
non-radial stellar pulsations. {\tt LP-PUL} provides the
eigenfrequency $\omega_{\ell, k}$ ---where $k$ is the radial order of
the mode--- and the dimensionless eigenfunctions $y_1, \cdots,
y_4$. {\tt LP-PUL} computes the periods ($\Pi_{\ell, k}$), rotation
splitting coefficients ($C_{\ell, k}$), oscillation kinetic energy
($K_{\ell, k}$) and weight functions ($W_{\ell, k}$). The expressions
to compute these quantities can be found in the Appendix of
\citet{2006A&A...454..863C}.  The Brunt-V\"ais\"al\"a frequency ($N$)
is computed as \citep{1990ApJS...72..335T}:

\begin{equation}
N^2 =\frac{g^2 \rho}{p}\frac{\chi_{\rm T}}{\chi_{\rm \rho}}(\nabla_{\rm ad}-\nabla + B),
\label{eq:brunt}
\end{equation}

\noindent where

\begin{equation}
B= -\frac{1}{\chi_{\rm T}}\sum_{i=1}^{n-1} \chi_{ X_{\rm i}} \frac{d \ln X_{\rm i}}{d \ln \rho},
\label{eq:ledoux}
\end{equation}

\noindent is the Ledoux term and contains the contribution 
coming from the chemical composition changes, and

\begin{equation}
\chi_{\rm T} = \left [\frac{\partial \ln p}{\partial \ln T} \right ]_{\rho}\quad~ 
\chi_{\rho} = \left [\frac{\partial \ln p}{\partial \ln \rho} \right ]_{\rm T}\quad~ 
\chi_{X_i} = \left [\frac{\partial \ln p}{\partial \ln X_i} 
\right]_{\rho, {\rm T}, X_{j \neq i}}\quad
\end{equation}

In what follows we describe how we generated the synthetic
  random-period spectra by means of a uniform distribution
  function. Even though the observed periods of real ZZ Ceti stars do
  not come from a uniform distribution --in fact they actually show a
  set pattern-- this is not an issue, because we use a quality function
  that self-selects the representative periods of ZZ Ceti stars. This
  will be further detailed in section
  \ref{sect:synthetic}. Particularly, with our method, we quickly
  generate a huge number of synthetic period spectra. They are used to
  perform a statistical approach in order to assess the impact of the
  uncertainties coming from the thermal pulses phase on the
  asteroseismologically-derived stellar parameters. Finally this
approach allows us to carry out full exploration of the impact on the
stellar parameters of ZZ Ceti models representative of stars from the
blue and red edge of the instability strip separately.

\subsection{ Synthetic set of periods}
\label{sect:synthetic}
In our first analysis we concentrated on asteroseismological fits
performed to sets of random synthetic periods. We generated 1000 
sets of three random periods (i.e. 1000  "artificial pulsating stars"), representative
of ZZ Ceti stars, from the \$RANDOM function in BASH. \$RANDOM returns
a pseudo-random integer with a periodicity of 
$16 \times (2^{31}-1)$. Then, we performed asteroseismological fits
to those artificial stars and compared the respective stellar
parameters that were obtained.

As shown by \citet{2006ApJ...640..956M}, those observed pulsating WDs
located at the blue edge of the ZZ Ceti's instability strip are
characterized by pulsation modes with periods < 350 s with small
amplitudes, whilst those located at the red edge are characterized by
pulsation modes with periods > 650 s and larger amplitudes. Motivated
by this findings, we performed
the asteroseismological fits and the corresponding  analyses for the
following period ranges: $\Pi_i <$ 350 s (RI), 350 $<\Pi_i <$ 650 s
(RII) and $\Pi_i >$ 650 s (RIII).

To find our best fit model we search for the one that best matches the
pulsation periods of our artificial stars. To do this, we calculate a
quality function as in \citet{2008MNRAS.385..430C}, and seek for the
model that minimizes it:

\begin{equation}
\phi=\phi(M_{\star},M_{\rm H},T_{\rm eff})= \frac{1}{N}\sqrt{\sum_{i=1}^N\frac{[\Pi_i^{th}-\Pi_i^{obs}]^2A_i}{\sum_{i=1}^N A_i}}
\label{eq:phi}
\end{equation}

\noindent
where the amplitudes A$_i$ are used as weights of each observed
period. Here, we used our artificial periods as if they were observed
periods and fixed the amplitudes as $A_i= 1$.  Since generally our
  quality function leads to very similar results as compared with those
  coming from a quality function that does not take into account the
  mode amplitude, we shall describe the quality of our period fits in
  terms of $\phi$ \citep{2012MNRAS.420.1462R}.
We set a threshold value of the quality function, $\phi_{lim}$, and
disregard those models characterized by $\phi > \phi_{lim}$.  For
artificial stars with periods in RI, we set $\phi_{lim}= 6$ s, while
for artificial stars with periods in the ranges RII and RIII, we set
$\phi_{lim}= 3$ s. These limit values are motivated from a balance
  between the amount of possible seismical solutions that satisfies
  the condition $\phi<\phi_{lim}$ and the smallest possible difference
  between the periods corresponding to the model and those
  characterizing the synthetic period spectra. 
 Out of the complete set of models, we find only 153 "best-fit models"
in the range RI, 293 in RII and 620 in RIII.

\begin{figure}

 \includegraphics[width=9.cm, angle=0]{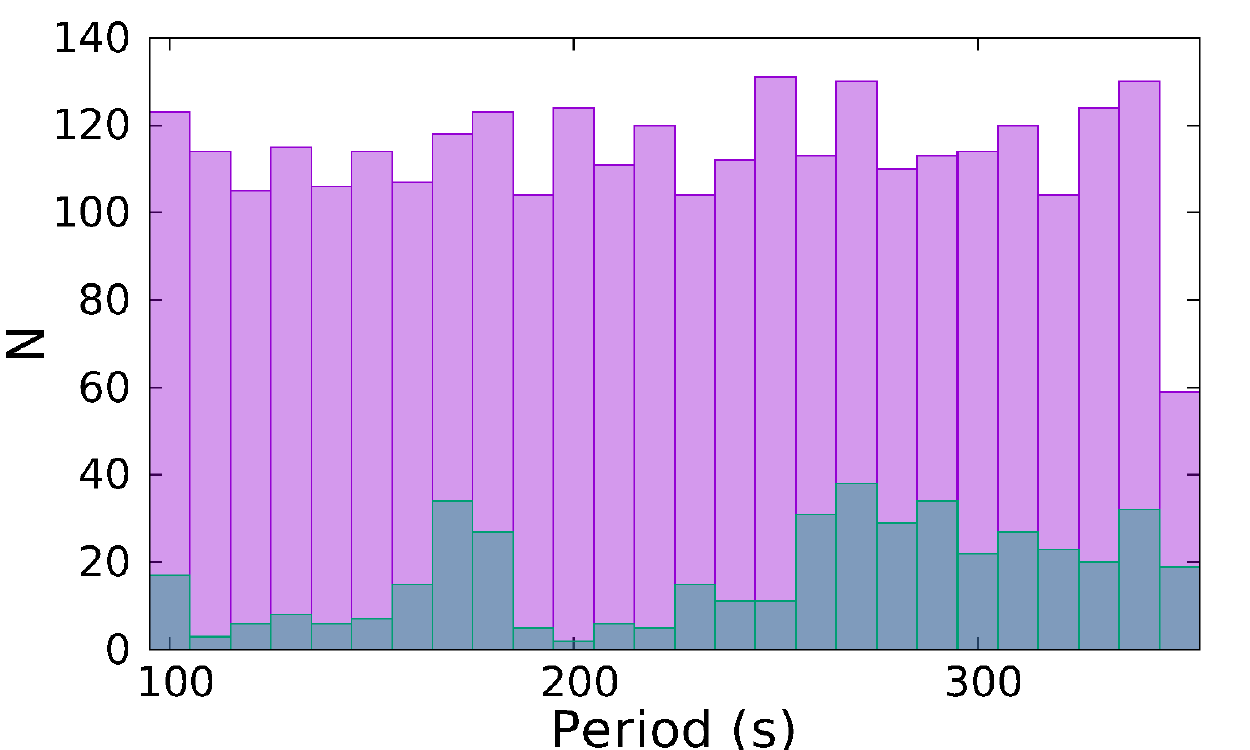}

 \caption{Histograms of the 3000 random periods in RI (pink bars) and
   of those of the 153 artificial stars which satisfy $\phi < 6$
   (effective periods: green bars).}
  \label{fig:periodos-hist-95-260}
\end{figure}

 At the outset, our synthetic periods were generated randomly with a uniform
distribution. However, by virtue of the constraint $\phi < \phi_{lim}$,
the effective periods ---those which correspond to a best-fit model--- 
are clustered at around 100, 170, 270 and 340 s, as depicted in 
Fig.\ref{fig:periodos-hist-95-260}.
This is in line with the findings of
\citet{2016arXiv161102579C} that, in hot ZZ Ceti stars, the pulsation 
periods are arranged in groups separated by gaps.
Specifically, by comparing the observational periods with those from
theoretical evolutionary computations of \citet{2012MNRAS.420.1462R},
\citet{2016arXiv161102579C} find that the existence of these gaps is
in agreement with models with a thick hydrogen layer mass.  In addition,
\citet{2016arXiv161102579C} compared period groupings with those
obtained by performing Monte Carlo simulations using the period grid
from \citet{2012MNRAS.420.1462R}  with different assumptions, 
and conclude that most of the hot ZZ Ceti have $M_{\rm H}$ values at or near the
canonical limit with He layers thinner than those predicted by
evolutionary models. As we will see later, this result is in agreement
with what we found. These period patterns are not found when studying
the remaining ranges of periods RII and RIII.

\section{Models and asteroseismological fits}
\label{cap:models-asteroseismological}

In this section we study the impact of the occurrence of TPs during
the progenitor evolution on the parameters of ZZ Ceti stars derived by
means of asteroseismology. As we mentioned, despite the fact
that the occurrence of the TP-AGB phase is expected for most of WD
progenitors, there is diverse evidence that some WDs could have
resulted from progenitor stars that avoided this phase. In this
context, as shown in
\citet{2017A&A...599A..21D}, whether the progenitor star evolves through the
TP-AGB or not strongly impacts the period spectrum of pulsating DA
WDs, specially in the case of the low-mass ZZ Ceti stars.

In this line, we developed a grid of evolutionary models with progenitor masses in the range $0.85 \leq M_{\rm ZAMS}/M_{\odot} \leq 2.25$ (final masses in the range $0.5349 \lesssim M_{\rm wd}/M_{\odot} \lesssim 0.6463$) 
from the ZAMS to the TP-AGB phase. On the AGB, we forced the evolutionary models to abandon this stage by 
enhancing the stellar mass loss rate at two
different stages: previous to the onset of the first thermal pulse, 
and at the end of the third thermal pulse (0TP and 3TP models,
respectively). As a result of the pulse driven convection zone that
develops at each TP, a  double-layered chemical structure with
an intershell region rich in C and He at the bottom of the He
buffer is expected \citep{2010ApJ...717..897A}.  
The size of this intershell and the precise shape of the chemical profile characterizing the
double layered region depend on the number of TPs experienced by the
progenitor as it evolves along the TP-AGB phase. In particular, as a
consequence of the outward moving He-burning shell, the size of the
intershell decreases and the core/He chemical transition shifts to
outer layers. Nevertheless, it is during the first TP that the main
chemical features of the intershell and the double layered regions
emerge. In the case of low mass stars, we do not expect the occurrence
of a large number of TP, so considering 3 TP is enough for capturing the essential of the chemical structure arising from TP-AGB phase. 
For higher stellar masses, we expect a larger number of  TPs
to take place, but as shown in \citet{2017A&A...599A..21D}, the period
spectrum expected at the ZZ Ceti stage is not markedly affected by the
number of additional TPs experienced by the progenitor star see their Fig. 3 (a, b). 

\begin{figure}
 \includegraphics[width=8cm, angle=0]{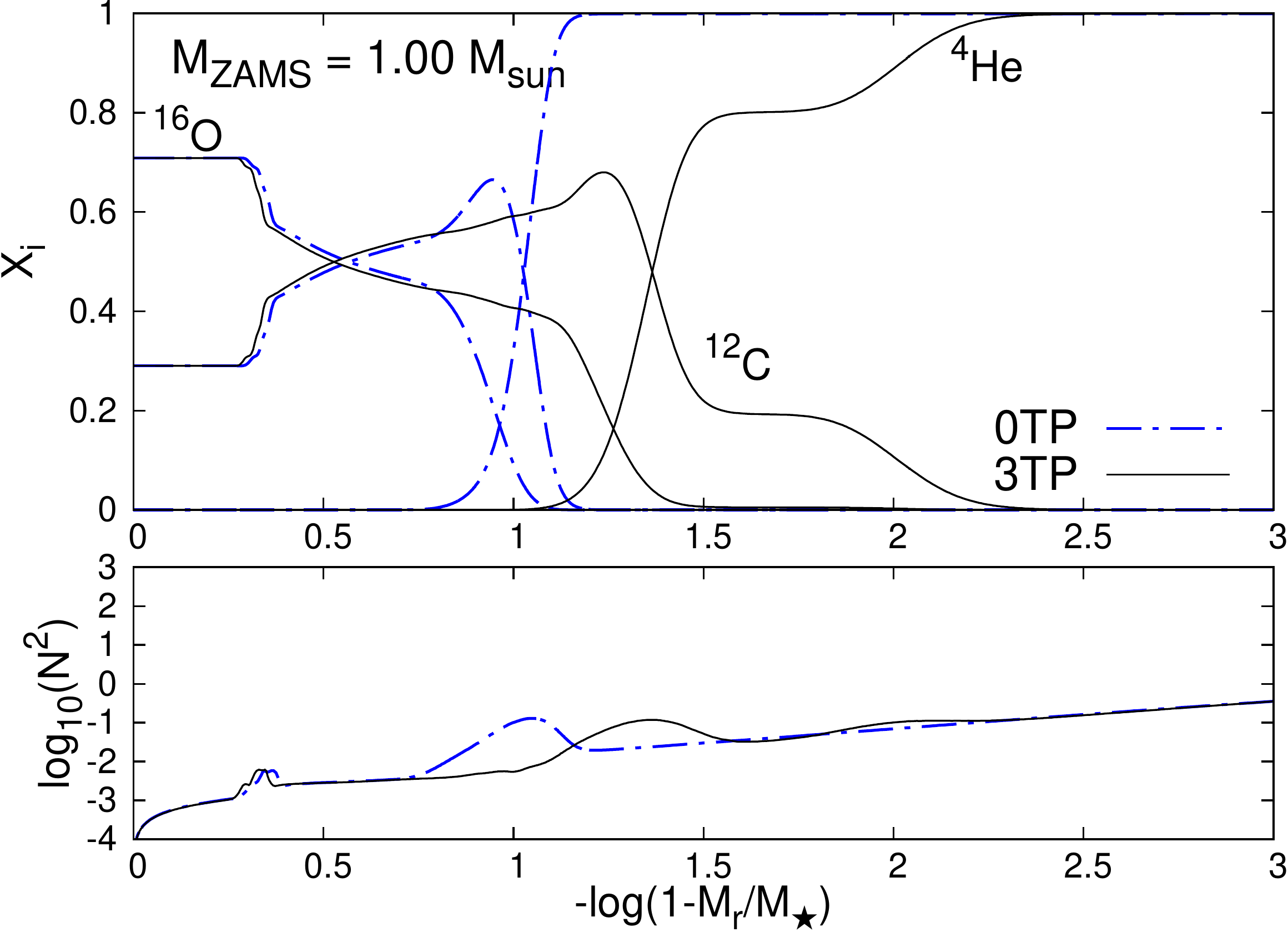}
 \caption{Internal chemical abundances for the most abundant species of
our 0TP and 3TP models for  a $\rm{}1 M_{\sun}$  progenitor at the ZZ Ceti stage
($M_{\rm wd}= 0.5508 M_{\odot}$). During the evolution through the first three TPs the core-He transition shifts from $-\rm{}\log(q) \sim 1$ to $\sim 1.3$ and the intershell region is built up.}
 \label{fig:perfil}
\end{figure}

Once the progenitor leaves the AGB phase, we continue its evolution to
the WD cooling phase until the ZZ Ceti stage,  where we calculate the
period spectrum.  In Fig. \ref{fig:perfil} we show the internal
chemical profiles for the most abundant elements of our models (upper
panel) and the logarithm of the squared Brunt-V\"ais\"al\"a (B-V) frequency (lower panel) in terms of the outer mass fraction for models with
$M_{\rm ZAMS}= 1.00 M_{\odot}$, at the ZZ Ceti stage ($T_{\rm eff} \sim 12\, 000$ K). For the 3TP model, three chemical transition regions from center to the surface can be noticed: an inner chemical interface of C and O, and intermediate interface rich in He, C, and O that separates the core from the intershell region rich in He and C, and finally an interface separating the intershell region from the pure helium buffer (an additional interface separating the pure H and He envelopes, not shown in the figure, is also present in the outermost layers). The bumps in the B-V frequency induced by these transition regions, strongly affect the pulsation spectrum and the mode trapping properties.  The imprints of the occurrence of TPs on the B-V frequency result from both the shift in the position of the core/He transition and the presence of the intershell region that emerges during the
TP-AGB phase.
As shown in \citet{2017A&A...599A..21D}, the intershell region
survives down to the ZZ Ceti stage only for low-mass WD stars. This is
because diffusion processes acting on the WD cooling phase are less
efficient for the less massive stars. On the other hand, for more
massive models, this intershell region is almost completely eroded by
the action of element diffusion, see Fig. 1 (b) of
\citet{2017A&A...599A..21D}. We mention that in our seismological fits
we have taken into account different values for the H content of the
WD. In particular, we considered the H envelope mass to vary from the
canonical value\footnote{ The value predicted from fully
    evolutionary computations.} to $\log (M_{\rm H}/M_{\rm wd}) \sim
-9$. 

To account for the differences induced in the stellar parameters of
asteroseismological models of ZZ Ceti stars resulting from considering
or not the TP-AGB phase, we performed period-to-period fits to a large
number of artificial stars that were generated by considering several
sets of three $\ell= 1$ mode periods. Based on the work of
\citet{2006ApJ...640..956M}, we explore the impact on artificial stars
belonging to different period ranges.  In order to find the best-fit
solution, we consider the quality function $\phi$
[Eq. (\ref{eq:phi})], and search for its minimum ---that is, the best
period-fit model-- in both 0TP and 3TP sets of models.  In general,
when asteroseismological fits are performed with real data, external
determinations such as effective temperature, stellar mass or surface
gravity can be used as filters to discard models that are in
disagreement with observational parameters, \citep[see][section
  4.3.1]{2012MNRAS.420.1462R}. In the case of artificial stars, we are
unable to use external constraints because the synthetic periods were
generated randomly without any assumption at the outset. In this case,
the results presented here reveal the differences between best-fit
models according to the lowest value of the quality function
only. 
In what follows, we compare the differences in the most relevant
 parameters of the asteroseismological models when they are derived from 
 the sets of 0TP and 3TP models.

\subsection{Impact over stellar parameters}

We employ a set of synthetic periods between 95 s and 350 s (typical
of hot ZZ Ceti stars), between 350 s and 650 s (typical of ZZ Ceti
stars in the middle of the instability strip), and between 650 s and
1250 s (typical of cool ZZ Ceti stars). Periods with values
  $\Pi_i \in$ [95 s -350 s] are associated to low-radial order modes
  and usually they probe the inner structure of the star. However, some
  of these modes could be sensitive to the outer layers of the star.

\begin{figure}

 \includegraphics[width=8cm, angle=0]{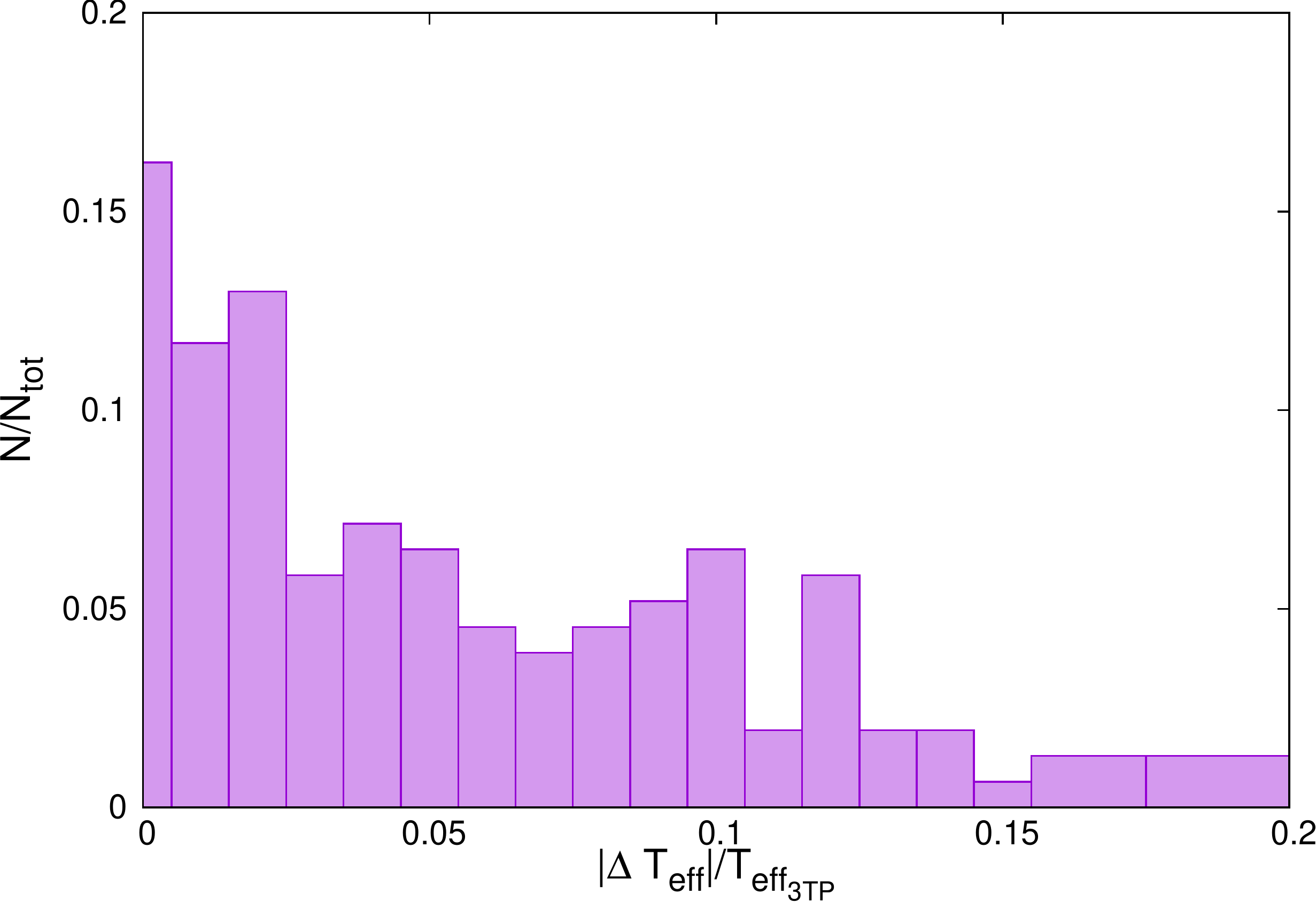}

 \caption{Histogram for the absolute differences in effective
   temperature between our best fits for the 0TP and 3TP sets of
   models in the $[95-350]$ s period range.}
  \label{fig:delta-teff-95-260}
\end{figure}

\begin{figure}
\includegraphics[width=9cm, angle=0]{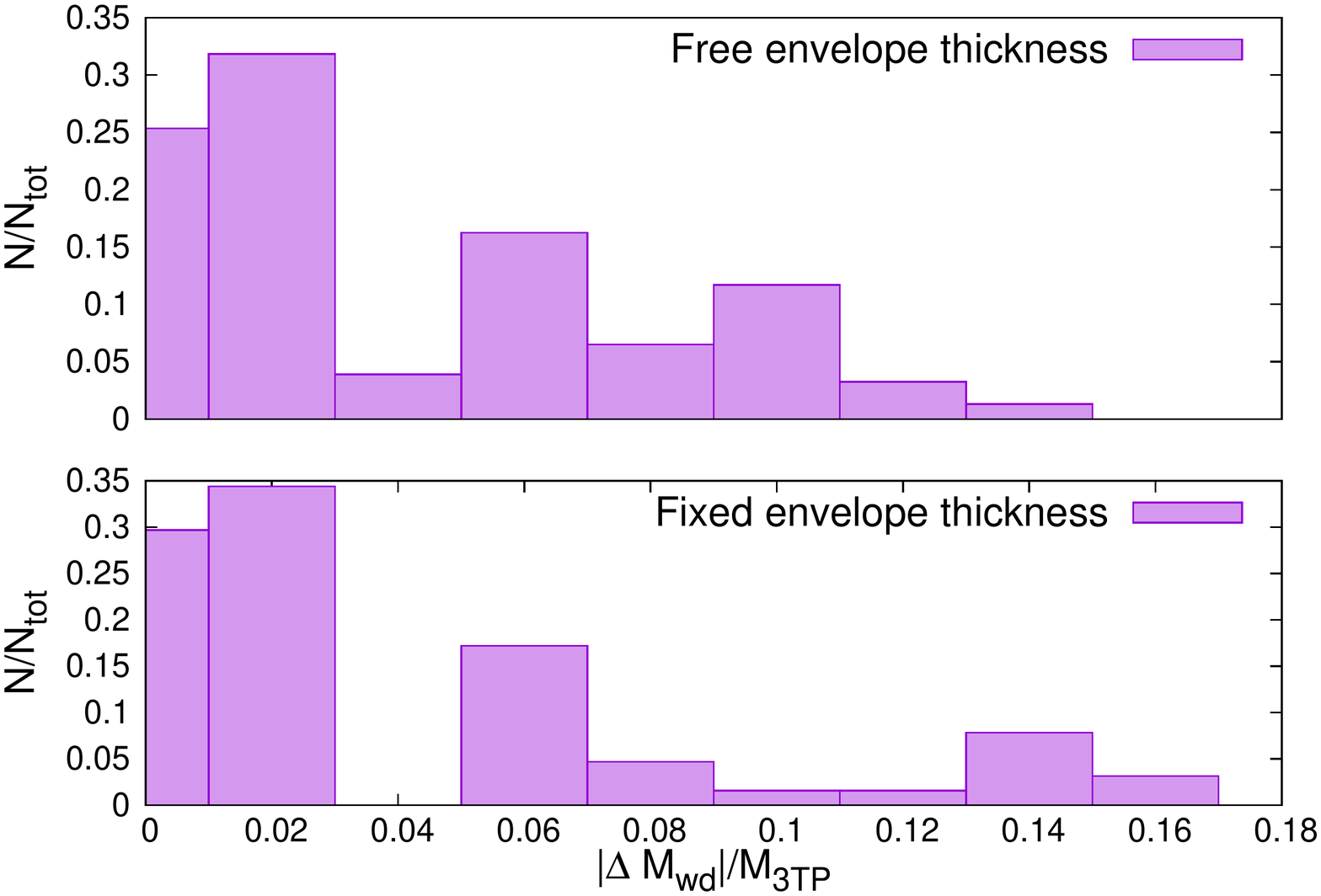}
   \caption{Histogram of the differences for differences in
   the stellar mass.}
  \label{fig:delta-masa-95-260}
\end{figure}

\begin{figure}
\includegraphics[width=8cm, angle=0]{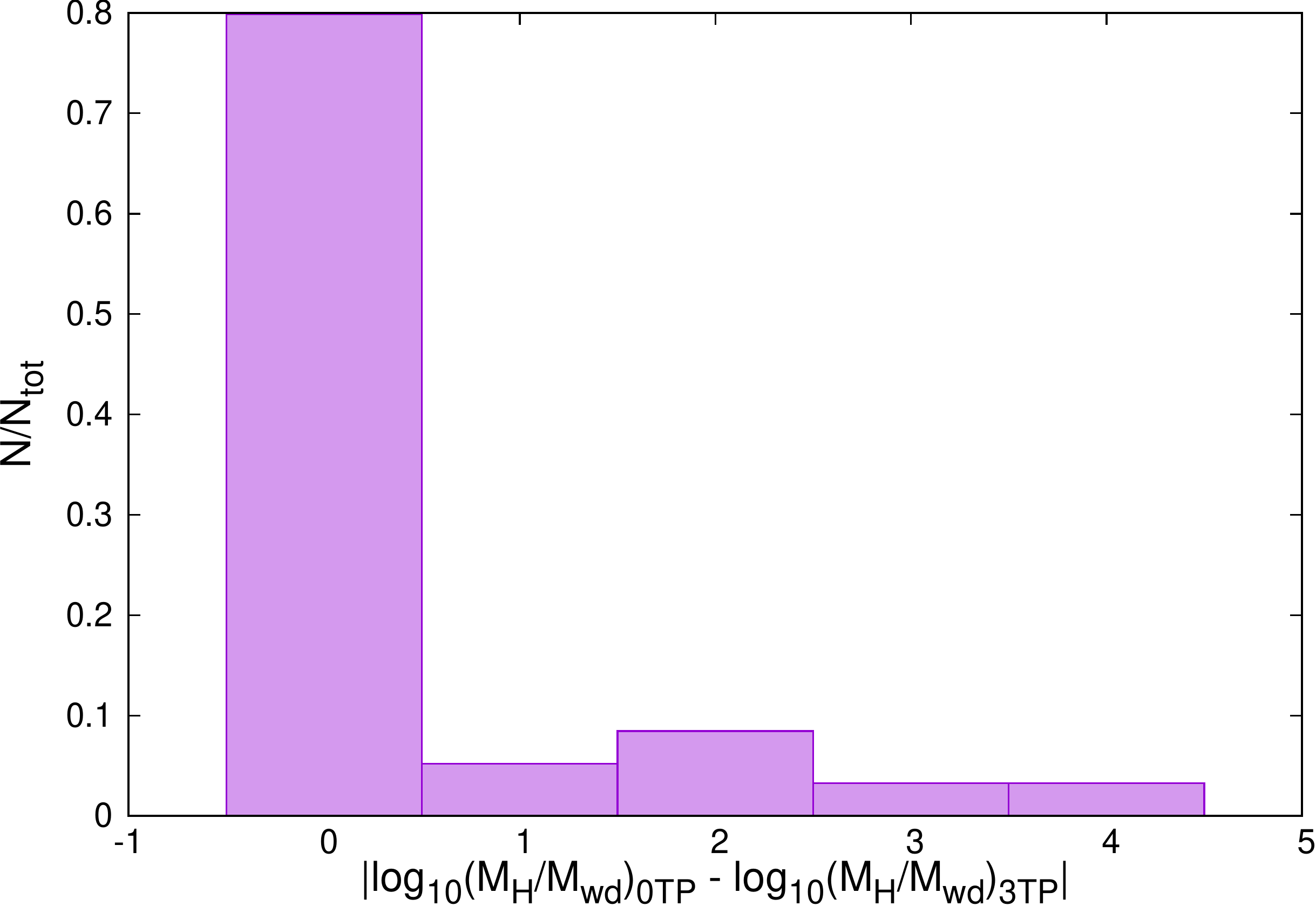}
\caption{Histogram for the differences in the determination of
  the H envelope mass for the period range [95-350] s. }
  \label{fig:delta-env-95-260}
\end{figure}

\begin{figure}

 \includegraphics[width=8cm, angle=0]{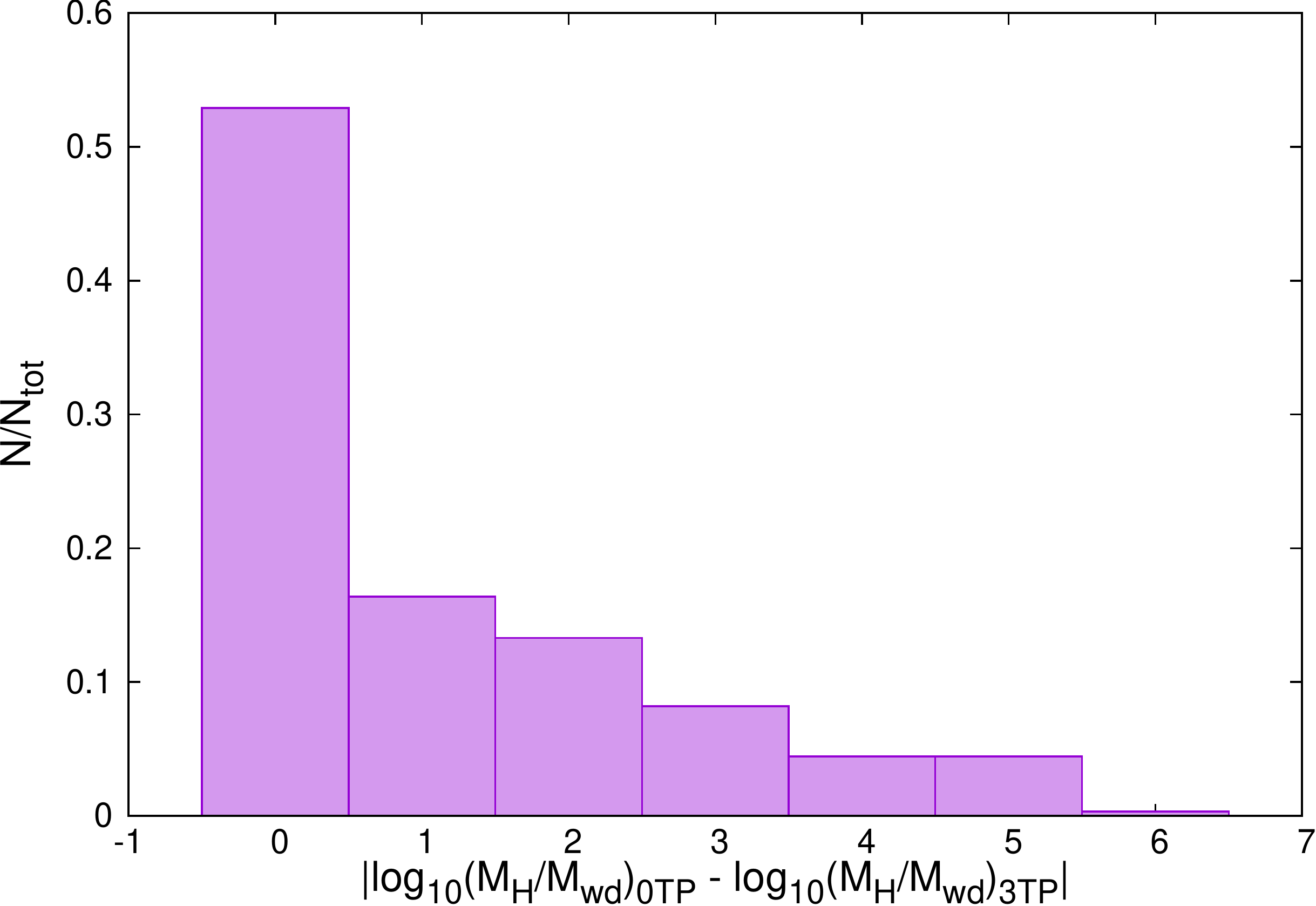}

 \caption{Same as Fig. \ref{fig:delta-env-95-260} but for fits considering
   the 350-650 s period range.}
  \label{fig:delta-env-350-300}
\end{figure}

\begin{figure}

 \includegraphics[width=8cm, angle=0]{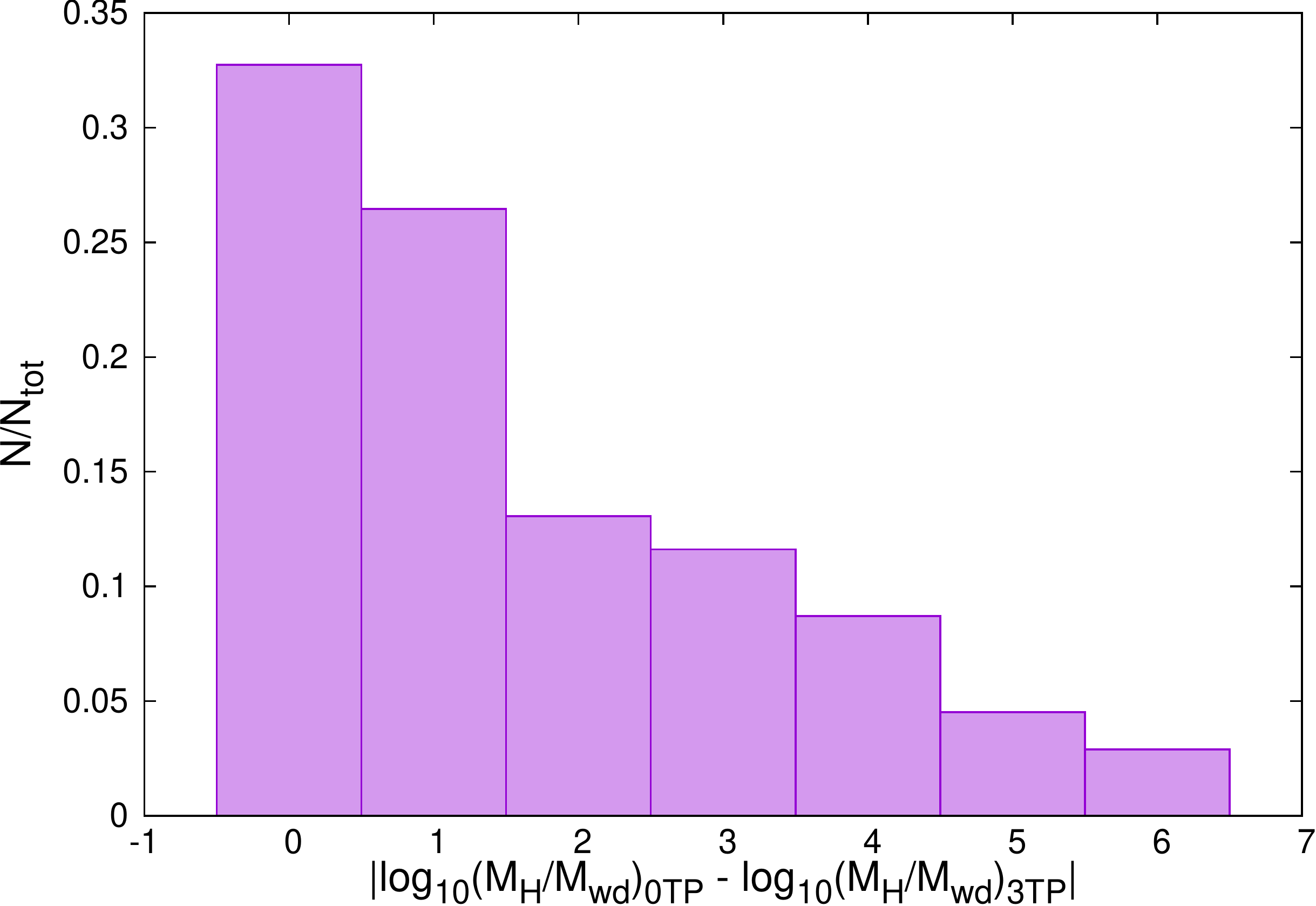}

 \caption{Same as Fig. \ref{fig:delta-env-95-260} but for fits considering
   the 650-1250 s period range.}
  \label{fig:delta-env-650-600}
  \end{figure}

In Figs. \ref{fig:delta-teff-95-260} to \ref{fig:delta-env-650-600} we
display histograms for the absolute difference in $T_{\rm eff}, M_{\rm
  wd}$ and $\log(M_{\rm H}/M_{\rm wd})$ between our 0TP and 3TP
best-fit models. These histograms represent the percentage ($N/N_{\rm
  Tot}$, $y$-axis) of artificial models with a given amount of
variation in the respective quantity\footnote{We take the models that
  experience three TP as our reference models.}  ($|\Delta T_{\rm
  eff}|/T_{\rm eff (\rm 3TP)}$, $|\Delta M_{\rm wd}|/M_{\rm wd (\rm
  3TP)}$, $M_{\rm H}/M_{\rm wd}$, $x$-axis).  In order to compare the
deviations found in the values of $M_{\rm H}/M_{\rm wd}$, we plot the
difference $|\log(M_{\rm H}/M_{\rm wd})_{\rm 0TP}-\log(M_{\rm
  H}/M_{\rm wd})_{\rm 3TP}|$.

From Fig. \ref{fig:delta-teff-95-260} we see that most of the relative
differences in effective temperature between both sets of fits for hot
ZZ Cetis are concentrated toward small values of $|\Delta T_{\rm
  eff}|$. A similar trend is also found for intermediate-$T_{\rm eff}$
and cool ZZ Ceti models.  We fit a Gaussian function of the form
$f(x)= A\cdot \exp(-(x-x_0)^2/(2\cdot\sigma^2))$ to the histogram and
obtain  
that the mean deviation in effective temperature
is roughly 700 K. The mean deviation is $\sim 500$ K and $\sim 800$ K for
intermediate-$T_{\rm eff}$ and cool ZZ Ceti stars, respectively. 
Although these mean variations may exceed the expected observational
uncertainties \citep{2011ApJ...730..128T}, it is worth noting that more than 50\%  
of the fits show differences lower than $\sim 300$ K.

Fig. \ref{fig:delta-masa-95-260} shows the expectations for the
differences in the stellar mass. In the upper panel, we show a
histogram of the differences that result when we allow to vary freely
all the parameters. This histogram displays several maxima at $|\Delta
M_{\rm wd}|/M_{\rm wd (\rm 3TP)} \sim 0.005, 0.02, 0.06, 0.11$.  For
cool ZZ Ceti models, we find that the number of cases showing larger
differences increases.
The occurrence of several maxima is probably linked to the well-known
core-envelope degeneracy related to the existence of a symmetry in the
mode-trapping properties which can lead to an ambiguity in determining
the location of features in the chemical structure \citep[see][for
details]{2003MNRAS.344..657M}.
To break this possible degeneracy, we redo all the asteroseismological
fits, but this time by considering a fixed H envelope mass of $\sim
10^{-6} M_{\rm wd}$ for all the models. The results are depicted
in the lower panel of Fig. \ref{fig:delta-masa-95-260}. Even though
some maximum still remain, about 70\% of the cases now show
differences smaller than the spectroscopic uncertainties, which can
rise up to 4-5\%\footnote{We mention here that when we fix the
  envelope mass, the effects on the distribution in $T_{\rm eff}$ are
  less pronnounced. In particular, the peak at $\sim 0.1$
  (Fig. \ref{fig:delta-teff-95-260}) is reduced and the number of the
  cases with deviations within 5\% increases.}. We emphasize
  that these differences in the stellar mass of the best fits models
  come from the fact that models with different stellar mass have
  different core chemical structure. For instance the extension in mass
  of the core and also the central chemical abundances of carbon and
  oxygen, depend on the stellar mass
  \citep{2010ApJ...717..897A,2012MNRAS.420.1462R}.

The distribution of the differences in $M_{\rm H}/M_{\rm wd}$
resulting from the two sets of evolutionary sequences is shown in the
histograms displayed in Fig.  \ref{fig:delta-env-95-260},
\ref{fig:delta-env-350-300}, and
\ref{fig:delta-env-650-600}. 
For the short period range, i.e. hot ZZ Ceti models (Fig. \ref{fig:delta-env-95-260}), most 
of the fits are clustered around $|\log(M_{\rm H}/M_{\rm wd})_{\rm 0TP}-\log(M_{\rm H}/M_{\rm wd})_{\rm 3TP}| \sim 0$, i.e., no differences in the H envelope mass are expected in most cases due to 
uncertainties in TP-AGB evolution. By contrast, as we consider cooler ZZ Ceti models,  
uncertainties in the TP-AGB phase impact markedly
the inferred H envelope mass (Figs. \ref{fig:delta-env-350-300}
and \ref{fig:delta-env-650-600}). This is not an unexpected behaviour, since some 
of the pulsational modes of high radial order characterizing cool ZZ Ceti stars are 
sensitive to the outer chemical structure. A Gaussian fit to the data yields a mean 
deviation of $M_{\rm H}$  
well below than an order of magnitude for hot and intermediate-$T_{\rm eff}$ models, and 2 order of magnitude for cool ZZ 
Ceti models. Clearly, for hot and intermediate ZZ Ceti models, the occurrence or not of TP during 
the AGB evolution of progenitor stars translates into uncertainties in the derived H envelope mass
of less than one order of magnitude in most cases. This is not the trend for the
case of cool ZZ Ceti models.

\subsection{Asteroseismological fits to real ZZ Ceti}

We contrast here the results of the previous section
with those derived from asteroseismological fits to some selected real
ZZ Ceti stars. We picked out those ZZ Ceti stars with modes previously identified
as $\ell=1$ 
and whose spectroscopic masses fall into the mass range considered in this
work. Spectroscopic determinations of $T_{\rm eff}$
and $M_{\rm wd}$ of the selected ZZ Ceti stars are shown in 
Table \ref{tab:prev-astero}. 
We classify these stars as cool, intermediate-$T_{\rm eff}$
or hot ZZ Ceti stars depending on the value of the pulsation 
period with the highest amplitude.  For the asteroseismological analysis 
and selection of the best fit models we have taken into account the 
criteria described in Sect. 4.3.1 of \citet{2012MNRAS.420.1462R}, i.e.:

\begin{itemize}
\item  the models minimize the quality function given by Eq. \ref{eq:phi}, 
so the observed periods are closely matched by the theoretical ones,
\item  we consider only those stars with assigned $\ell= 1$ modes in previous asteroseismological analysis,
\item we elect those models with $T_{\rm eff}$ and $\log(g)$
as close as possible to the spectroscopic determinations.
\end{itemize}

\begin{table*}[]
\centering

\begin{tabular}{llll}

\hline
\multicolumn{0}{l}{}
Star~~~~~~~~~~~~~~~~& $T_{\rm eff}$ & $M_{\rm wd}/M_{\odot}$ & Reference\\
\hline
KUV 11370$+$4222 & 11890   & 0.639                &\citet{2004ApJ...600..404B}\\

HE 0031$-$5525   & 11480  & 0.44                &\citet{2006AandA...450..227C}\\

WD J1002$+$5818  & 11710   &0.57                 &\citet{2005ApJ...625..966M}\\

WD J0214$-$0823  & 11570   &0.57                 &\citet{2004ApJ...607..982M}\\

BPM 31594      & 11450   & 0.666               & \citet{2004ApJ...600..404B}\\

G191$-$16        & 11420   & 0.632               &\citet{2004ApJ...600..404B}\\ 

MCT 0145$-$2211  & 11500   & 0.684               &\citet{2004ApJ...600..404B}\\

WD J150$-$0001  & 11200   & 0.61                &\citet{2004ApJ...607..982M}\\

HE 1429$-$037    & 11434   & 0.514               &\citet{2005AandA...443..195S}\\

EC 23487$-$2424  & 11520   & 0.661               &\citet{2004ApJ...600..404B}\\        

G232$-$38        & 11350   & 0.610               &\citet{2006AJ....132..831G}\\

\hline
\end{tabular}

\caption{Spectroscopic determinations of
  $T_{\rm eff}$ and $M_{\rm wd}/M_{\odot}$ corresponding to the ZZ Ceti
  stars selected in this work.}
\label{tab:prev-astero}

\end{table*}

\begin{figure*}[h]
\centering
\includegraphics[width=1.0\textwidth]{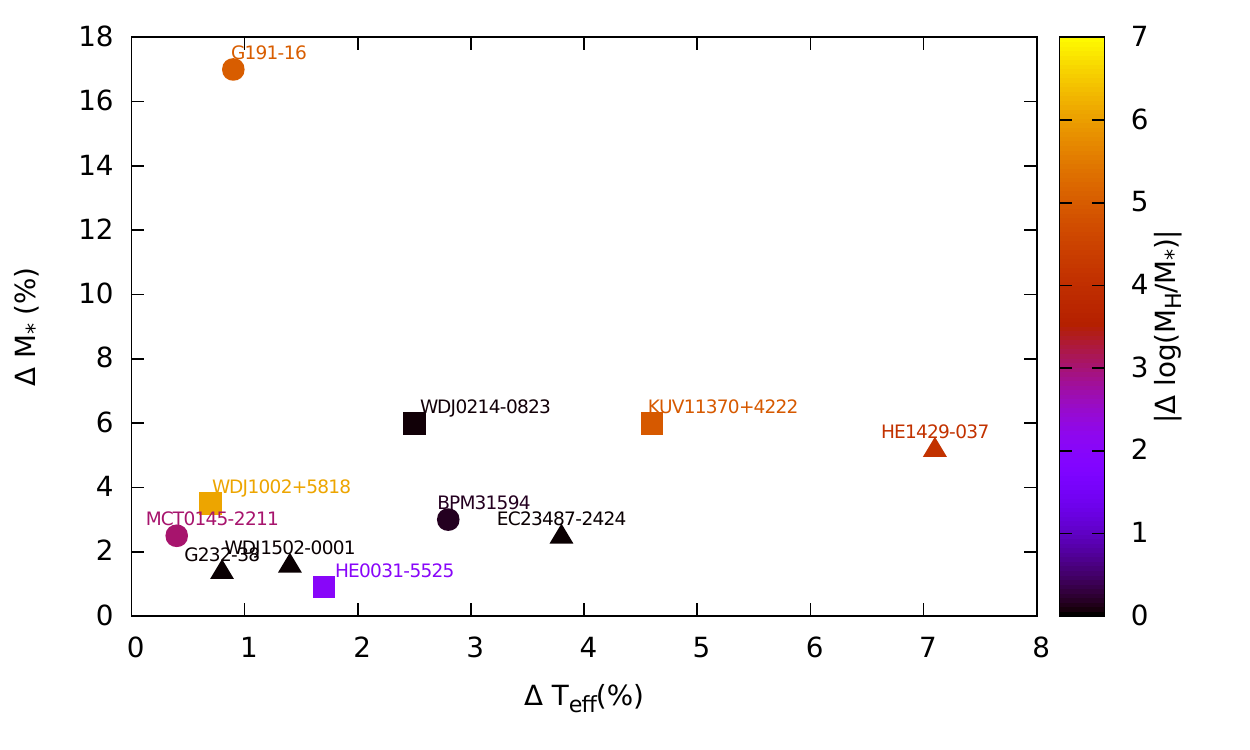}
\caption{Variations of  $T_{\rm eff}$, $M_{\rm wd}$  and $M_{\rm H}$ --in color scale-- of selected ZZ Ceti stars resulting from the two sets of evolutionary sequences considered in this work. Squares, circles, and triangles refer to hot, intermediate, and cool ZZ Ceti stars, respectively.}
\label{pt-real-zzceti}
\end{figure*}

The results of the asteroseismological period-to-period fits are
illustrated in Fig. \ref{pt-real-zzceti}, that shows the deviations
found in the derived stellar parameters as given by the two sets of
evolutionary sequences. In this figure we plot the percentage of the
deviation in effective temperature and stellar mass, together with the
absolute difference in $\log(M_{\rm H}/M_{\rm wd})$, as given by the
color scale to the
right. 
The differences in $T_{\rm eff}$ resulting from the occurrence of TPs
range from 0.4\% to 7.1\%. The differences in stellar mass are in the
range 0.9-6\%, except for G191$-$16, for which the difference in
stellar mass amounts to 17\%.  
 Although we might expect a certain link between the magnitude of
  the deviations found on the different stellar
  parameters, 
  we did not found a clear pattern. Another important result from
Fig. \ref{pt-real-zzceti} is that, independently of the periods
exhibited by the stars, a wide range of colors is seen, i. e., some
hot- and even intermediate-effective temperature ZZ Ceti stars are
sensitive to the thickness of the hydrogen
envelope. 
 There were some hints before that even low radial overtone modes
  could be sensitive to the structure in the outer layers of the
  models. For instance \citet{2012MNRAS.420.1462R} found that a low
  $k$ mode in the hot ZZ Ceti G117-B15A was very sensitive to the mass
  of the hydrogen envelope.

\section{Summary and conclusions}

\label{cap:conclusions}

In this work, we have studied the impact of the occurrence of TPs
during the AGB evolution of the WD progenitor stars on the stellar
parameters of ZZ Ceti stars as derived from asteroseismological period
fits to artificial and real ZZ Ceti stars. To this end, we evolved
progenitor star models with initial masses in the range 0.85 to 2.25
$M_{\odot}$ from the ZAMS to the TP-AGB phase. Here, we forced the
progenitor to abandon the AGB at two stages: previous to the
occurrence of the first TP and at the end of the third TP. In this
way, we generated two sets of evolutionary sequences, 0TP and 3TP. All
of the resulting sequences were evolved along the WD stage down to the
domain of the ZZ Ceti stars, at $T_{\rm eff} \sim 12\,500-10\,500$ K.
At this stage, we computed the theoretical period spectrum for each
model.  Asteroseismological period-to-period fits were carried out to
a set of random periods as well as to the period spectra of selected
real ZZ Ceti stars.

Marked differences in the pulsational period spectrum are expected
depending on whether the progenitor star evolves or not through the
TP-AGB phase. We find that this translates into a non-negligible
impact for the resulting asteroseismological fits, depending on the
period range exhibited by the star. We report that the fact that a WD
progenitor experiences or not the TP-AGB phase implies an average
deviation of the effective temperature of the asteroseismological
models for ZZ Ceti stars of at most 8\% (with more than 50\% of the
fits showing differences lower than 300 K), and of the order of
$\lesssim$ 5\% in the stellar mass (with more than 70\% of the fits
showing differences lower than the spectroscopic uncertainties). For
the mass of the H envelope, however, we find deviations up to 2 order
of magnitude in the case of cool ZZ Ceti models. For hot and
intermediate ZZ Ceti models, no differences in the H envelope mass is
expected in most cases from the occurrence of TP during AGB evolution.
These trends remain when real ZZ Ceti stars are considered. We
  also find that the period spectrum of some hot and intermediate
  effective temperature ZZ Ceti stars are sensitive to the thickness
  of the hydrogen envelope (see Fig. \ref{pt-real-zzceti}).

 This paper, which is part of an
ongoing project, assesses for the first time the uncertainties in
progenitor evolution ---particularly the occurrence of the TP-AGB
phase--- and their impact on stellar parameters inferred from
asteroseismological period fits to ZZ Ceti stars.  Even though the
TP-AGB phase is expected to be a common phase for most of WD
progenitors, evidence is mounting that some WDs could have been formed
from progenitor stars that avoided this phase; for instance, this
would be the case for most of the He rich stars in NGC2808, which fail
to reach the AGB phase, thus evolving directly to the WD state after
the end of the He core burning \citep{2017arXiv170602278M}. As shown
in \cite{2017A&A...599A..21D} the occurrence or not of TPs is the main
uncertainty in progenitor evolution that mostly affects the expected
pulsational periods of ZZ Ceti stars. As far as asteroseismology is
concerned, the results found in this paper show that the imprints left
by the occurrence of the TPs during AGB evolution of progenitor star
are not negligible and must be taken into account in
asteroseismological fits of these stars.

As stated before, two main asteroseismological avenues are
being currently applied to peering into the interior of WDs,
both methods being complementary to each other. On one hand,
there is the approach that considers stellar models with 
parametrized chemical composition profiles. This is a powerful
method that has the flexibility of allowing a full exploration of
the parameter space to find an optimal asteroseismological model.
The downside of this method is that sometimes it can lead to WD
asteroseismological models with chemical structures
that are not predicted by stellar evolution (e.g., the existence of a
pure C buffer, unrealistic abundances of C and O at the core,
etc).  The second approach, developed at La Plata Observatory,
is different in nature, as it employs fully evolutionary models
that are the result of the complete evolution of the progenitor stars,
from the zero-age main sequence (ZAMS) until the WD phase. 
This method involves the most detailed and updated input physics,
in particular regarding the internal chemical structure expected
from the nuclear burning history of the progenitor, a crucial
aspect for correctly disentangle the information encoded in the
pulsation patterns of variable DA WDs. However, we emphasize 
that this method is affected by several important uncertainties connected with
evolutionary processes during the progenitor star evolution.
A specific assessment of the impact of these uncertainties
on the properties of asteroseismological models of ZZ Ceti stars
derived with this method was lacking. This paper just come to fill this gap.
Specifically, in this paper and in \citet{2017A&A...599A..21D} we demonstrate that the uncertainties
in prior WD evolution do affect WD asteroseismology,
but that the effects are quantifiable and bounded. Indeed,
differences in stellar mass, effective temperature and H envelope
thickness due to the occurrence or not
of TP at the AGB phase, the main uncertainty resulting from the evolutionary history of progenitor star,  are within the typical
spectroscopic errors. These results add confidence to the use
of fully evolutionary models with consistent chemical profiles, and
render much more robust our asteroseismological approach.
In order to complete this critical evaluation of our
asteroseismological method, we plan to present a future paper 
that addresses the impact of the uncertainties in the
$^{12}$C$(\alpha,\gamma)^{16}$O nuclear
reaction rate on the asteroseismological inferences of ZZ Ceti stars
(De Geronimo et al. 2018, in preparation).

\begin{acknowledgements}
 We acknowledge the valuable comments of our referee that
  substantially improved the original version of this paper.  We
warmly acknowledge the helpful comments of J. J. Hermes.  Part of this
work was supported by AGENCIA through the Programa de Modernizaci\'on
Tecnol\'ogica BID 1728/OC-AR and by the PIP 112-200801-00940 grant
from CONICET. This research has made use of NASA Astrophysics Data
System.
\end{acknowledgements}


\bibliographystyle{aa} 
\bibliography{paper-incertezas} 

\end{document}